# Is *Atomic Force Microscopy - Infrared* a Photothermal Technique?

*Luca Quaroni*

*Department of Physical Chemistry and Electrochemistry, Faculty of Chemistry, Jagiellonian University, 30-387, Cracow, Poland; luca.quaroni@uj.edu.pl*

## Abstract

Atomic Force Microscopy - Infrared (AFM-IR) has emerged as a technique for measuring absorption spectra with spatial resolution better than the optical diffraction limit. The technique relies on the movement of a probe for atomic force microscopy for detecting the local expansion of a material caused by the photothermal effect. While AFM-IR is seeing increased application to a wider range of samples, reports have also appeared in the literature that are inconsistent with an interpretation of the AFM-IR response simply in terms of photothermal expansion. The present perspective addresses the issue by critically evaluating existing experimental observations.

The photothermal effect converts absorbed light into heat, causing a local temperature increase in the absorbing region. The heating and the associated thermal expansion are wavelength dependent and track the absorption coefficient of a material, providing the foundation for several spectroscopic techniques. In one implementation, the deflection of a cantilevered tip has been used to monitor the thermal expansion at the tip-sample contact region.[1] The resulting spectroscopic technique, promises spatial resolution comparable to that of Atomic Force Microscopy (AFM), even when working with excitation in the mid-infrared (MIR) region, where the relatively long wavelengths ($\mu \sim 2.5 - 25$ μm) allow only a modest diffraction-limited resolution. The method has received different names, among which, AFM-IR (Atomic Force Microscopy – Infrared) is more widespread and will be used in this text.

The assumption underlying AFM-IR is that the cantilever movement is a monotonic function of the temperature increase and the photothermal expansion, and can be fully described by the thermoelastic properties of probe and sample material and by the absorption coefficient of the latter.[2–4] However, analysis of existing reports suggests that the mechanism of signal generation in AFM-IR could be more complex than expected, comprising contributions from forces other than the ones associated to the photothermal expansion.

The basic principle of photothermal spectroscopy is that the signal must increase monotonically with the local temperature increase. With a modulated or pulsed light source, the temperature distribution corresponds to the distribution of the resulting thermal wave, a heavily damped local oscillation of temperature that satisfies the diffusion wave formalism.[5] The spatial distribution of the corresponding photothermal expansion must track the spatial distribution of the temperature increase (Figure 1A). The extension of a thermal wave is dependent on the modulation frequency of the light source and on the thermal diffusivity of the absorbing medium, and is quantified by its thermal diffusion length, $L$, (Equation 1), or by the maximum extent of the temperature perturbation after one excitation cycle, $R$ (Equation 2), which can be used as measure of resolution.[6] *f* is the modulation frequency of the excitation source and *α* is the thermal diffusivity of the material. Higher frequencies probe smaller and shallower regions, improving both lateral resolution and surface selectivity (Figure 1A). Common values for *R* and *L* are in the approx. range 0.1 µm – 1.0 µm, thus supporting the claims of resolution better than the optical diffraction limit.

$$L = \sqrt{\frac{\alpha}{\pi f}} \text{ (Equation 1)} \quad R = \sqrt{\frac{3\alpha}{f}} \text{ (Equation 2)}$$

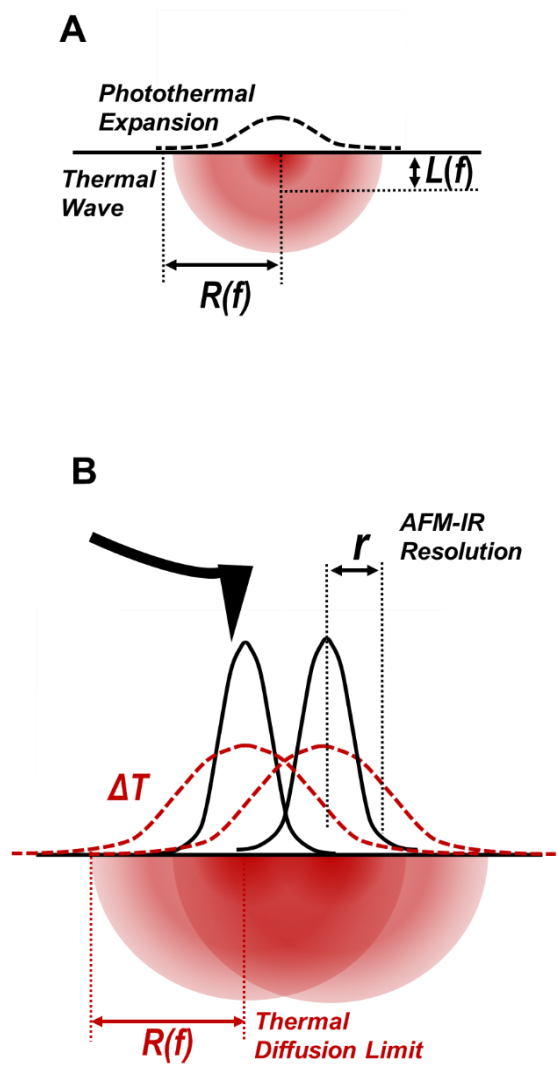

*Figure 1. Performance Parameters of a Photothermal Measurement.* **A.** *Resolution in the thermal diffusion limit (R) and thermal diffusion length (L) of photothermal measurements are dictated by the size of the thermal wave formed by the absorption of light modulated at frequency, f. The extension of the thermal wave defines the region subject to thermal expansion.* **B.** *In contrast, in AFM-IR measurements the movement of the cantilever responds to a smaller region than defined by the thermal wave, providing better resolution than predicted by thermal diffusion.*

In contrast to theory, the resolution claimed in several AFM-IR publications is better than expected from thermal wave distribution. A clear discrepancy between measured AFM-IR resolution and temperature increase has been noted in comparative STIRM (Scanning Transmission Infrared Microscopy) measurements.[7] STIRM images confirm that the size of the region of temperature increase is of the order of a few hundred nanometers, in qualitative agreement with multiple predictions [2,4,6] The resulting temperature profiles follow the spatial profile of thermal waves, as calculated by Mandelis *et al.*.[8] However, parallel AFM-IR images on the same samples do not track the temperature increase.[4] (Figure 1B). The discrepancy has been confirmed by the comparison of theoretical and measured photothermal resolution versus pulsing frequency.[9] Measured resolution improves with increasing frequency and approaches the contact radius of the tip at higher frequency, in agreement with a photothermal origin of the signal. However, the frequency dependence is modest, and measured values are about one order of magnitude lower than expected from values of *R*.[9] Overall, AFM-IR provides better spatial resolution than predicted by the distribution of the associated thermal waves.

The analysis of signal intensity also reveals discrepancies. For example, calculations by Morozovska *et al.* [4] predict a monotonic, but not linear, increase of the AFM-IR signal from a supported thin film. In contrast, calculations by Mandelis *et al.* predict a decrease of the thermal wave amplitude with increasing thickness, an effect caused by thermal wave interference within the film. [8] Measurements by Lahiri *et al.* show an initial linear increase of the AFM-IR signal at lower thickness, followed by a decrease at higher thickness. While the latter effect may be connected to the bottom-up illumination geometry, the inconsistencies also suggest multiple overlapping effects.[10]

Following these observations, I propose that, while AFM-IR appears to respond to the impulse from photothermal expansion, other forces can also contribute to the movement of the probe. In addition to Van der Waals forces, already accounted for in the description of tip-sample contact mechanics, other possible contributions are electrostatic and magnetic forces, photoinduced optical gradient forces, and photoacoustic forces.[11–15] Among them, the photoinduced force (PiF) is the basis of the technique for nanoscale spectroscopy that bears its name, which provides a spatial resolution much better than allowed by thermal waves. Except for the photoacoustic force, these additional forces arise from interactions between localized charges or dipoles in the sample and a polarizable tip apex, with a dependence from tip - sample distance, d, according to laws of the form $F \propto 1/d^n$, where n is an integer that depends on the type of interaction and the geometry of the system. The inverse power laws provide a rapid decay with distance that is consistent with the high spatial resolution reported experimentally. The effect of these forces can manifest itself when the tip-sample distance is varied during the measurement, whereby they can modulate the amplitude of the cantilever oscillation, but it is absent when contact is maintained. Therefore, displacement of the tip from the sample must be occurring to explain deviations from an ideal photothermal behavior in terms of both intensity and resolution.

Modulation of the tip-sample distance is inherent to the operation of tapping mode AFM-IR. However, current theoretical descriptions of contact mode AFM-IR assume that contact between tip and sample is retained throughout the measurement[2,3]. Consequently the expansion directly excites supported (contact) modes of the cantilever, producing oscillations with a nodal point at the tip-sample contact. Despite its widespread adoption, this hypothesis has not been confirmed or challenged by targeted experiments. I now propose that, even when operating in contact mode, loss of tip-sample contact and excitation of unsupported (non-contact) modes must occur to account for the observed deviations from the expected photothermal response. Possible causes include the impulse delivered by the expanding sample and/or the deflection of the cantilever itself caused by heating.[16,17]

The overlapping contributions of multiple forces pose a limit to the quantitative analysis of complex samples by AFM-IR. As an example, the presence of localized charges or dipoles can affect signal intensity and image contrast, modulating the response independently from compositional factors. On the positive side, these interactions also constitute an opportunity for contrast generation and for analyzing additional properties of the sample. Identifying non-photothermal contributions to the AFM-IR signal will be a challenge for the coming years. Possible approaches include measurement of the dependence of the signal from the tip sample distance, as done by O'Callahan *et al.* for a non-contact configuration,[15] and/or comparison with the response of reference samples of known properties under standard conditions. Identification of all contributing forces should be a necessary preparatory step for experiments that claim quantitative accuracy. There is more to AFM-IR than the photothermal effect, and it should not be neglected.

## Acknowledgements

Writing up of this work was in part supported by an OPU16 grant from the National Science Center Poland under contract UMO-2018/31/B/NZ1/01345.